\documentclass[%
 preprint,
 amsmath,amssymb,
 aps,
]{revtex4-1}

\usepackage{graphicx}
\usepackage{hyperref}
\usepackage{dcolumn}
\usepackage{bm}

\usepackage{indentfirst}

\begin{document}

\preprint{}

\title{River-bed armoring as a granular segregation phenomenon}

\author{Behrooz Ferdowsi}

\altaffiliation[equal contributions]{}
\affiliation{%
 Earth and Environmental Science, University of Pennsylvania, \\Philadelphia, PA 19104, USA.
}%
\author{Carlos P. Ortiz}%

\altaffiliation[equal contributions]{}
\affiliation{%
 Earth and Environmental Science and Department of Physics and Astronomy, University of Pennsylvania,\\ Philadelphia,
PA 19104, USA.}%

\author{Morgane Houssais}%
\affiliation{%
Benjamin Levich Institute,\\ The City College of New York, New York, NY 10031, USA.}%

\author{Douglas J. Jerolmack}%
\altaffiliation[corresponding author]{}
\affiliation{%
Earth and Environmental Science, University of Pennsylvania,\\ Philadelphia,
PA 19104, USA.}%
 \email{sediment@sas.upenn.edu}

\date{\today}

\begin{abstract}
\noindent
Gravel-river beds typically have an ``armored'' layer of coarse grains on the surface, which acts to protect finer particles underneath from erosion. River bed-load transport is a kind of dense granular flow, and such flows are known to vertically segregate grains. The contribution of granular physics to river-bed armoring, however, has not been investigated. Here we examine these connections in a laboratory river with bimodal sediment size, by tracking the motion of particles from the surface to deep inside the bed, and find that armor develops by two distinct mechanisms. Bed-load transport in the near-surface layer drives rapid segregation, with a vertical advection rate proportional to the granular shear rate. Creeping grains beneath the bed-load layer give rise to slow but persistent segregation, which is diffusion dominated and insensitive to shear rate. We verify these findings with a continuum phenomenological model and discrete element method simulations. Our results suggest that river beds armor by granular segregation from below --- rather than fluid-driven sorting from above --- while also providing new insights on the mechanics of segregation that are relevant to a wide range of granular flows.
\begin{description}

\item[Keywords]
granular flow, creep, riverbed armoring, granular segregation
\end{description}
\end{abstract}

\maketitle


\section*{\label{sec:level1}Introduction}
\noindent
River-bed grain size controls the exchange of solutes, nutrients and fine particulates across the sediment-fluid interface \citep{packman2004hyporheic, brunke1997ecological}, and determines the flood magnitude required to initiate motion {\citep{phillips2014dynamics, mueller2005variation, wilcock1993critical}. Grain size, however, also evolves over a series of floods as particles are sorted longitudinally and vertically during transport  {\citep{parker1982gravel,dietrich1989sediment,fedele2007similarity, phillips2014dynamics}. A ubiquitous pattern observed in gravel-bed rivers is armoring, in which the median grain size of the surface is significantly larger than that of the subsurface (Fig.~\ref{fig1} A). Laboratory experiments designed to simulate gravel rivers --- i.e., bed-load transport of heterogeneous grain sizes --- reproduce the phenomenon, but disagree on its origins. Three competing mechanisms have been proposed: (i) kinetic sieving, in which smaller particles migrate downward through the void spaces between larger particles during motion \citep{frey2009river}; (ii) 'equal mobility', whereby the proportion of large and small surficial grains adjusts to achieve a spatially constant entrainment stress \citep{parker1982gravel}; and (iii) sediment supply imbalance, in which the transport capacity of the flow locally exceeds the upstream supply and results in surface coarsening \citep{dietrich1989sediment}. All of them assume that gravel in transport only mixes with the substrate over a small ``active layer'' that is one to several grain diameters deep.

Recently, sediment transport experiments have revealed that granular motion extends much deeper into the subsurface {\citep{aussillous2013investigation,houssais2015onset}. In particular, grains transition continuously from rapid bed-load motion at the surface to slow creeping motion far below the surface \citep{houssais2015onset,houssais2015rheology}. Both kinds of motion also occur in dry granular systems, where bed-load corresponds to a dense granular flow, and creep is characteristic of quasi-static deformation of disordered granular packs \citep{houssais2015onset,houssais2015rheology}. The former is known to produce robust vertical size segregation by kinetic sieving {\citep{savage1988,makse1997possible,gray2005theory}. Phenomenological continuum models based on this premise  \citep{gray2015particle,gray2005theory,fan2011theory,staron2015stress} produce vertical segregation that is consistent with experimental observations \citep{gray2015particle,wiederseiner2011experimental} and discrete element method (DEM) simulations \citep{fan2014modelling,hill2014segregation}}. Segregation by creep is unexplored; while reports of slow coarsening do exist \cite{Finger2014}, its connection to creep has not been demonstrated. 

\begin{figure*}[ht]
\centerline{\includegraphics[width=1.0\textwidth]{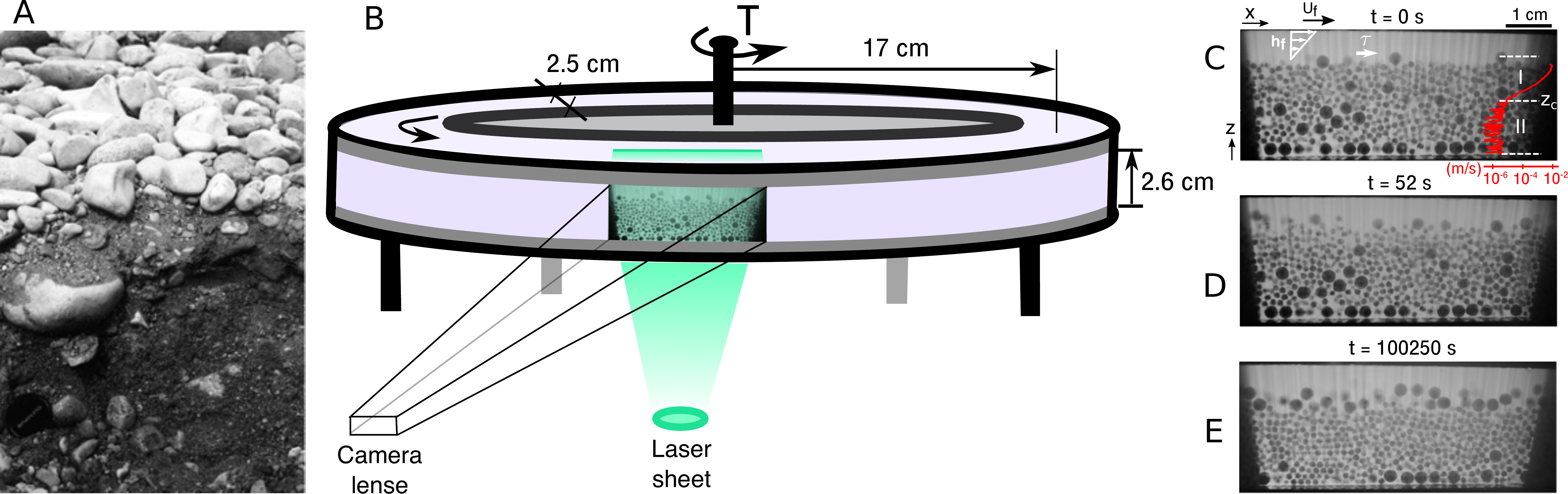}}
\caption{Phenomenology and setup. (A) Bed sediment of the River Wharfe, U.K., that shows a pronounced surface armor. Photo courtesy D. Powell \citep{powell1998patterns}. (B) Sketch of the experiment, showing position of the camera and laser plane used for imaging inside the granular bed. (C-E) Snapshots during armor development for $\tau_s^* = 3.8\tau^*_{cs}$. Also shown is the fluid boundary stress, which is computed as $\tau =  \eta U_f/h_f$ \citep{houssais2015onset} where $U_f$  and $h_f$ are the top-plate speed and flow depth, respectively. The red curve shows the long-term-averaged streamwise particle velocity $u_x(z)$, where I and II correspond to the bed load and creep zones, respectively. The directions $x$ and $z$ are indicated.}\label{fig1}
\end{figure*}

The contribution of granular physics to river-bed armoring has only begun to be examined. Frey and Church \citep{frey2009river} showed with laboratory experiments that bed load drives segregation by kinetic sieving that is qualitatively similar to dense granular flows. Here we investigate granular segregation and quantify its contribution to armoring using an idealized laboratory river experiment. Our setup [Fig.\ref{fig1}B] is designed to: eliminate the disruptive influence of flume boundaries by using an annulus; image particle motion from the sediment-fluid interface to deep in the subsurface, away from the wall; isolate granular contributions by simplifying particles to bimodal spheres and eliminating fluid turbulence with a viscous fluid; and explore a range of transport conditions from near threshold to vigorous bed load. These experiments demonstrate how river-bed armor can develop due to bottom-up motion of subsurface grains, while revealing new insight on granular segregation mechanisms in a system where rapid and slow granular flows co-exist. Results are compared with predictions from a modified phenomenological segregation model, and with DEM simulations of a dry-granular bed under shear.

\section*{Experiments}
\noindent
Experiments were conducted in a closed-top annular flume (Fig.\ref{fig1}B); details of the apparatus have been described previously \citep{houssais2015onset,houssais2015rheology}. The channel walls are smooth to allow slip between grains and the boundary, in order to approximate an infinitely deep and wide channel. The flume is filled with a bidisperse granular bed of acrylic spherical grains with small and large diameters, $d_{s}=1.5$ mm and $d_{l}=3.0$ mm, respectively, and density $\rho_p=1.19$ $g$ $\text{mL}^{-1}$; the ratio of total small to large grain volume in the channel is $\frac{V_{small}}{V_{large}} = 2$. The grains are submerged in a fluid of viscosity $\eta = 72.2$ mPa s and density $\rho=1.05$ $g$ mL${}^{-1}$. A fluid gap is sheared from above by rotating the lid of the flume to apply a constant fluid-boundary shear stress, $\tau$ (Fig.\ref{fig1}C); it is reported here as dimensionless Shields number for the small grains, defined as $\tau_{s}^*=\frac{\tau}{(\rho_p-\rho) gd_{s}}$, where  $g$ is gravity. The associated Shields stress for large grains as $\tau_{l}^* = \frac{d_s}{d_l} \tau_{s}^*$. For reference, Shields numbers for each experiment are compared to the critical Shields number, $\tau^*_c$, that is classically used to identify the onset of sediment transport. Our previously determined critical Shields number for a monodisperse bed of small grains, $\tau^*_{cs} \simeq 0.1$ \citep{houssais2015onset}, is used here as the reference critical stress, recognizing that the actual value may differ in this bidisperse system \citep{houssais2012bedload}. We determined empirically for the present experiments that the range of Shields numbers $\tau^*_{cs} \le \tau^*_s \le 5\,\tau^*_{cs}$ corresponds to bed-load transport: a thin surface layer of moving grains in frequent contact with, and supported by, an underlying granular bed that is creeping \citep{houssais2015onset}. We report data from experiments conducted at five Shields numbers, $\tau_s^* = [2.7, 3.8, 4.1, 4.4, 4.7] \tau^*_{cs}$. All flows were laminar (Reynolds number $\le 4$) and grain collisions were viscously damped (Stokes number $< 1$) (see SI, section 1). 

The bed at the start of each experiment was composed of sedimented particles forming an approximately flat granular bed (see Methods). At the beginning of an experiment ($t=0$ $s$) fluid shear was initiated at the specified Shields stress, and applied for a duration of $24\,\text{hr}$ or longer. We image a cross-section of particles from the bed surface ($z_s$) to the bottom of the channel through time using refractive-index matched scanning \citep{dijksman2012invited} (Figs.~\ref{fig1}B; S2). Vertical profiles of streamwise particle velocity ($u_x(z)$) for experiments at all Shields stresses were determined from averaging pixel strips in the streamwise direction over all time, using image cross-correlation (SI, section 4). Velocity profiles confirm the existence of two distinct regions of particle motion (Fig,~\ref{fig1}C). Zone I corresponds to bed load, where velocity decays rapidly with depth below the surface; below this is zone II associated with creep, characterized by a much slower decay \citep{houssais2015onset}. All runs show a qualitatively similar evolution of the bed through time: a coarse surface ``armor'' layer develops as large grains are delivered from below; first more rapidly by bed load, and then more slowly by creep (Fig.~\ref{fig1}, SI Movies \href{https://www.dropbox.com/s/nabr47jripqmhem/bidisperse_realtime.mp4?dl=0}{1} \& \href{https://www.dropbox.com/s/4x844uurydrk7bv/bidisperse_timelapse.mp4?dl=0}{2}). This is explored in more detail below.

\begin{figure*}[ht]
\begin{center}
\centerline{\includegraphics[width=1.0\textwidth]{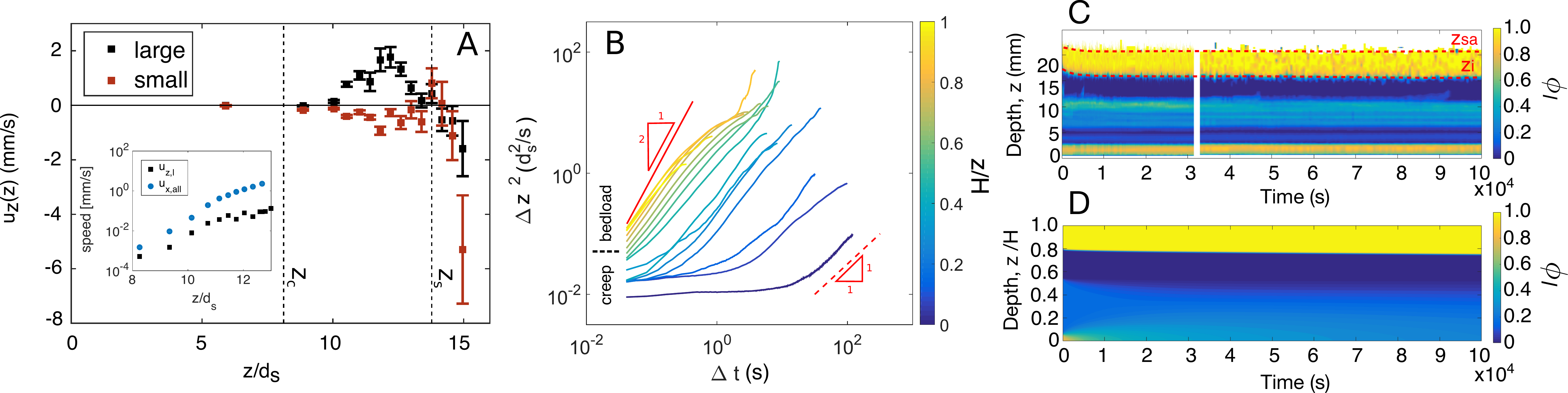}}
\caption{Experimental particle and segregation dynamics. (A) Vertical velocity profile for small and large grains for the interval $\Delta t=[0:20]$ minutes at the beginning of shearing at Shields number $\tau_s^* = 4.1\tau^*_{cs}$. The elevations of the bed surface ($z_s$) and transition to creep ($z_c$) are indicated. Inset shows the horizontal streamwise velocity $u_{x}$ profile for all grains and vertical velocity for large grains $u_{z,l}$ in logarithmic scale for the bedload zone. The streamwise particle velocity measurements have error-bar (one standard deviation/mean) value $\sim 0.3\%$, whereas vertical particle velocity measurements have error-bar (one standard deviation/mean) value $\sim 1\%$. (B) Vertical mean-square displacement (MSD) for large grains as a function of time $\Delta t$. MSD shown for various depths of the granular bed, defined with a colorbar. The top and bottom red-dashed lines indicate the limiting behaviors of advection and diffusion, respectively. Boundary between bed load and creep is indicated. Note that near-surface grains are advective at short times; creeping grains show no change in MSD at short times indicating caged dynamics, but transition to diffusive behavior at longer times. (C) 1D ($x$-averaged) concentration map of large grains over time for shear stress $\tau_s^* = 4.1\tau^*_{cs}$. The red dashed lines show the positions of the armor surface ($z_sa$) and the bottom of the armored layer ($z_i$), which are used to calculate the thickness of the armored layer in Figure \ref{fig3} A. (D) Same as (C) for the advection-diffusion model, using velocity profiles and initial conditions that correspond to the shear stress $\tau_s^* = 4.1\tau^*_{cs}$ in (A).}
\label{fig2}
\end{center}
\end{figure*}

In order to probe the size- and depth-dependent behavior of grain motion, and its contribution to vertical segregation, we construct trajectories of all imaged grains using the particle tracking method \citep{houssais2015onset} (see Methods) for a representative experiment at Shields stress $\tau_s^* = 4.1\,\tau^*_{cs}$. Profiles of average vertical velocity for large ($u_{z,l}$) and small ($u_{z,s}$) grains, computed from these trajectories, show a striking pattern: they are anti-correlated in the bed-load regime, with net upward (positive) velocity for large grains and net downward (negative) velocity for small grains. Although there are deviations in the near-surface (within $1 d_s$ of $z_s$) due to intermittent saltation, below this region the velocity of large grains decreases with depth and reaches approximately zero at the transition from bed load to creep ($z_c$) (Fig. \ref{fig2}A). The decay rate of $u_{z,l}$ is roughly exponential, and coincides with the decay of the bulk streamwise velocity $u_{x}$ (Fig. \ref{fig2}A-inset). This suggests that the observed vertical advection of larger grains is linked to horizontal granular shear in the bed-load zone.

Grains in the creep zone have a small but detectable vertical velocity. To determine the dominant modes of particle motion in bed load and creep, we inspect the scaling of the mean-square displacements (MSD) versus time. For the same experiment at $\tau_s^* = 4.1\,\tau^*_{cs}$ we compute the vertical MSD as a function of depth for the large grains as $\text{MSD}(\Delta t) \equiv \Delta z^2 (t) = \langle \big\vert z(t+\Delta t) - z(t)\big\vert^2\rangle$ over a duration of 20 minutes; the brackets indicate ensemble averaging over grains and the reference time $t$, and $z$ is the particle's vertical position (Fig.~\ref{fig2}; see Methods). A distinction can be made between grains above and below the depth associated with the transition from bed load to creep. Grains in the bed-load zone exhibit MSD growth at short times that approaches ballistic motion, and is consistent with the advection described earlier (Fig.~\ref{fig2}A). The strength of the advection behavior diminishes at larger timescales where it perhaps transitions to super-diffusive behavior. In contrast, grains in the creep zone appear to exhibit caged dynamics in which MSD grows slowly or not at all at short timescales. Motion transitions toward diffusive and sometimes super-diffusive dynamics at longer times. The crossover timescale indicates the average lifetime of cages, and it increases with depth into the creeping zone. This behavior is similar to what has been observed in slow granular flows \citep{wandersman2012particle, marty2005subdiffusion, choi2004diffusion}, and indicates that particle movement in creep is related to creation and destruction of the granular contact network \citep{howell1999stress,choi2004diffusion}.

To visualize the resulting development of surface armor, we examine the spatio-temporal concentration map of large grains; $\phi_l$ represents the streamwise-averaged areal fraction at a given depth and time (see Methods). The development of surface armor is seen as a high-concentration surficial layer that thickens through time (Figs.~\ref{fig2}C; S4-S8). We quantify the thickness of the armor layer (Fig.~\ref{fig2} C) as  $z_{sa} - z_{i}$, where  $z_{sa}$ and $z_{i}$ are the position of the top and bottom surfaces that define the armor layer, respectively (SI, section 5) for all five Shields stress experiments. The data suggest the existence of two stages in the creation of armor, anticipated by the granular dynamics described above (Fig.~\ref{fig3} A). First is rapid segregation (duration of $10^2$ - $10^3 s$), as large grains are delivered up from the shallow subsurface. The rate of segregation shows a strong dependence on the driving Shields number, consistent with shear-rate dependent segregation of bed load. Once the bed-load zone is depleted of large grains, there follows a slow but persistent segregation that continues for the duration of the run ($\sim 24\, \text{hr}$). We interpret the slow stage of segregation as creep driven. Interestingly, the rate of segregation in this stage is insensitive to the driving Shields number, suggesting that creep segregation does not depend strongly on the driving shear rate. 

Armor development in our experiments results from a vertical flux ($z$ direction) of coarse grains toward the bed surface. We quantify this segregation flux, $J$, as the time derivative of the number of large grains in the armored layer:

\begin{equation} \label{eq:sg0}
J = A \frac{d}{dt} \int_{z_i}^{z_s} \phi_l dz
\end{equation}

where $A$ is the cross-sectional area of the armor interface in the $x-y$ plane. The variation of segregation flux density ($J/A$) with time (Fig.~\ref{fig3}B) clearly shows the existence of two stages of armor formation. We introduce a dimensionless time $t/t_{adv}$, where the characteristic advection time $t_{adv} = \frac{h_{bl}} {\langle u_{z,l} \rangle} \sim \frac{h_{bl}} {aU_{sf}}$; $\langle u_{z,l} \rangle$, $U_{sf}$ and $h_{bl}$ are the average large grain velocity, the average surficial grain velocity and the thickness of the bed-load layer, respectively, and $a = \langle u_{z,l}\rangle / U_{sf} \sim 10^{-3}$ is measured for the experiment at $\tau_s^* = 4.1\,\tau^*_{cs}$}, but the value for $a$ collapses all data later in Fig.~\ref{fig3}C. We also define a dimensionless segregation flux $J/J(0)$ where $J(0)$ is the initial value for $J$ at the start of each experiment. Utilizing the dimensionless time and flux variables produces a reasonable collapse of the data (Fig.~\ref{fig3}C). For all experiments $J/J(0)$ decays to a value of $1/e$ at a characteristic dimensionless time of $O(1)$.}

\begin{figure*}[htp]
\begin{center}
\centerline{\includegraphics[width=1.0\textwidth]{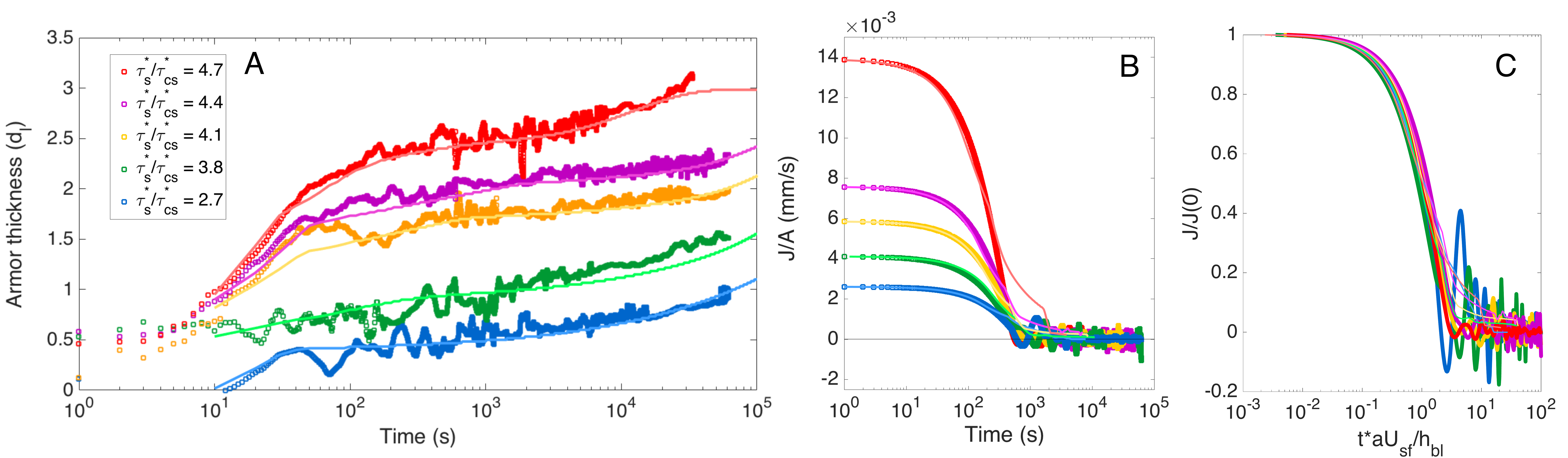}}
\caption{Armor thickness and segregation flux through time. (A) Temporal evolution of the thickness of the armored layer at different Shields number. Legend indicates the Shields number associated with each curve, and applies to (B) and (C) also. The brighter continuous lines are predictions from the advection-diffusion model Eq. \ref{eq:sg1}. Note the first rapid stage of armoring which is dependent on Shields number and is associated with bed-load transport, and the second slower stage that exhibits a nearly constant rate for all Shields numbers and is the result of creep. (B) The variation of segregation flux density ($J/A$) with time. (C) Normalized flux against dimensionless time; data are reasonably collapsed.}\label{fig3}
\end{center}
\end{figure*}

\section*{Advection-diffusion segregation model}
\noindent
Sediment transport produces armoring that appears similar to reported granular segregation experiments \citep{gray2015particle,savage1988}, implying that the presence of a viscous fluid has little influence beyond determining the shear rate of surficial grains. In particular, some previous experiments in dry granular flows suggested that segregation rate depends on the granular shear rate \citep{golick2009mixing, may2010shear}, consistent with our findings for bed load (although another study found otherwise \citep{wiederseiner2011experimental}). In addition, a recent study found that particle diffusion was shear-rate dependent for rapid granular flows but independent of shear rate for creep \citep{fan2015shear}, similar to our experiments. Because the exact mechanism of segregation is still a subject of debate \citep{staron2015stress,hill2014segregation,schroter2012granular}, there is no universally agreed upon continuum theory. Nonetheless, one-dimensional (1D) continuum models generally describe the vertical evolution of concentrations of binary mixtures through time with a phenomenological advection-diffusion equation \citep{gray2015particle,fan2014modelling}. Here we develop and apply a modified version of one such model, the Gray-Thornton model \citep{Gray2015,gray2005theory}. The model requires specification of: vertical advection and diffusion coefficients, usually assumed to be constant \citep{gray2005theory}; the vertical granular velocity profile; and the initial concentration profile. It then solves for the temporal segregation of large and small grains subject to mass conservation constraints. Two new ingredients must be included to account for the granular dynamics observed in our experiments: (i) for the bed-load regime, both advection and diffusion depend on shear rate; and (ii) for the creep regime there is no advection, and diffusion is independent of shear rate \citep{fan2015shear}. Our modified advection-diffusion segregation model, written in terms of the evolution of the large-grain concentration $\phi_l$, becomes:

\begin{equation} \label{eq:sg1}
\frac{\partial \phi_l}{\partial \hat{t}} + \frac{\partial}{\partial \hat{z}}\left(S_{rn}(\hat{z}) F(\phi_l)\right) = \frac{\partial}{\partial \hat{z}} \left(D_{rn}(\hat{z}) \frac{\partial \phi_l}{\partial \hat{z}}\right).
\end{equation}

Equation \ref{eq:sg1} is written in terms of dimensionless elevation $\hat{z} = z/H$ and time $\hat{t} = tU_{sf}/L$, where $H$ and $L$ are the height of the granular pack and the length of the centerline of the annular flume. The flux function $F(\phi_l)$ determines the dependence of the segregation flux ($S_r F(\phi_l)$) on $\phi_l$. Although there are ongoing debates on the mathematical form of the flux function \citep{van2015underlying,Gray2015}, we implement the simplest choice: a quadratic function $F(\phi_l) = \phi_l(1-\phi_l)$ that is symmetric about $\phi_l = 0.5$, which assumes that small and large grains behave identically but in opposite directions. The original Gray-Thornton model assumed a non-dimensional advective segregation velocity $S_r$ that is independent of shear rate. We introduce a depth-dependent parameter, $S_{rn}$, in order to redistribute the non-dimensional advective segregation velocity, $S_r$, according to the depth-dependent grain velocity, $u_x(\hat{z})$, normalized by the vertical average of grain velocities, ${\langle u_x(\hat{z}) \rangle}$  (Eq. \ref{eq:sg2}). 

\begin{equation} \label{eq:sg2}
S_{rn}(\hat{z}) = \left\{
  \begin{array}{lr}
    S_r   \frac{\beta \exp(\beta \hat{z})}{\exp(\beta)-1} & : \hat{z} \ge \hat{z}_c\\      
    0 & : \hat{z} < \hat{z}_c
  \end{array}
\right.
\end{equation}

The form of the normalized velocity is determined by a fit to the bed-load velocity profile such that:

\begin{equation} \label{eq:sg4}
\frac{u_x(\hat{z})}{\langle u_x(\hat{z}) \rangle} = \frac{\beta \exp(\beta \hat{z})}{\exp(\beta)-1}
\end{equation}

where $\beta$ is the exponential decay constant of the bed-load velocity profile.  Accordingly, for our analysis we define the parameter $S_r = \frac{L}{H{\langle u_x(\hat{z}) \rangle}}q$, where $q$ is the maximum bulk advective segregation velocity, i.e., that associated with the start of the experiment ($t = 0$; see SI section 6; Fig. S3). Similarly, we introduce a dimensionless and vertically-varying diffusivity $D_{rn}$ that has the same exponential decay as the velocity profile characterized by $\beta$ (Eq. \ref{eq:sg3}). The parameter $D_r = \frac{DL}{H^2{\langle u_x(\hat{z}) \rangle}}$ is a non-dimensional diffusive-remixing constant, where $D$ is the dimensional diffusivity.

\begin{equation} \label{eq:sg3}
D_{rn}(\hat{z}) = \left\{
  \begin{array}{lr}
    D_r \frac{ \exp(\beta \hat{z})}{\exp(\beta)-1} & : \hat{z} \ge \hat{z}_c\\
    D_r \frac{ \exp(\beta \hat{z})}{\exp(\beta)-1} & : \hat{z} < \hat{z}_c
  \end{array}
\right.
\end{equation}

To apply the new model Eq. \ref{eq:sg1} to our experiments requires specification of several parameters, determined from each experimental run (see Methods and SI). The input velocity profile $u_x(z)$ is determined by fitting two exponential functions to the time-averaged velocity profiles of the bed-load and creep zones, respectively (see Methods; Fig. S6). The input value for $q$ is computed as the upward migration velocity of the center of mass for the large particles at the start of each experiment (see Methods). Note that the advective segregation term in Eq. \ref{eq:sg2} decays with decreasing velocity (and depth) in the bed-load zone, and is set to zero in the creep zone ($z < z_c$). The diffusivity also decays with velocity (and depth) in the bed-load zone, and is constant for creep (Fig. S6 E). We take the dimensionless diffusion constant $D_r$ as a fitting parameter. In particular, the ratio $S_r/D_r$ is estimated by fitting the position of the armor interface through time for each experiment. We find that a constant ratio of $S_{rn}/D_{rn} \sim 318$ for the bed-load zone and $S_{rn}/D_{rn} =0 $ for the creep zone is sufficient to describe the development of armor for all Shields numbers.  We use the profile of $\phi_l$ at the start of our experiments ($t = 0$) as the initial concentration profile for the continuum model (Fig. S6F).

A visual comparison of armor development for the example condition $\tau_s^* = 4.1 \tau^*_{cs}$ shows that the advection-diffusion segregation model (Eq. \ref{eq:sg1}) captures the experimental behavior well (Fig.~\ref{fig2}D). A more quantitative comparison of the thickness of the armored layer through time (Fig.~\ref{fig3} A) demonstrates good agreement between the model and data for all Shields numbers. Importantly, the model correctly captures the initial fast and subsequent slow stages of segregation. The large ratio $S_{rn}/D_{rn} \sim 318$ for $z > z_c$ confirms the idea that the rapid stage of armor development is driven by shear-rate dependent advection associated with bed load. 
The fact that the ratio \(S_{rn}/D_{rn}\) remains constant for all experiments suggests that the model results are robust. 
The bulk kinetics can be related to particle-scale advection and diffusion by noting that $S_{rn}/D_{rn}=\frac{S_r}{D_r}\beta=Pe \frac{H}{d}\beta$, where the particle-scale ratio of advection to diffusion is given by the Peclet number \(Pe=u_z d_l/D_z \). For the experiment with \(\tau_s^* = 4.1 \tau^*_{cs}\) we determined from measurements that \(u_z=1.51\)~mm/s and \(D_z=3.38~\text{mm}^2\)/s, which leads to $Pe=1.3$ and $S_{rn}/D_{rn}=140$; the latter is the same order of magnitude as the ratio used in the continuum simulations.
The creeping zone is characterized by a constant value for $D_r$, and a lack of advection ($S_r = 0$), for $z < z_c$. This supports the notion that the slow stage of armoring results from diffusion by creeping grains that is independent of local shear rate.

\section*{Discrete Element Modeling}
\noindent
The analysis presented thus far shows how explicit accounting for the kinematics of granular motion in bed load and creep can produce a reasonable continuum description of armor development. In order to demonstrate that the observed armoring in experiments is entirely a consequence of granular physics, we now turn to DEM simulations in which the velocity profile and segregation dynamics arise spontaneously from grain-grain interactions. Simulations are performed with \href{http://www.cfdem.com/liggghtsr-open-source-discrete-element-method-particle-simulation-code}{LIGGGHTS}, an open-source granular modeling package based on LAMMPS (\url{http://lammps.sandia.gov}). Details of model implementation are available in Methods and Supplementary Information. In accord with the low Stokes number of our laboratory experiments, the restitution coefficient is chosen to be very small ($e_n = 0.01$) such that collisions are highly damped (Table S2). Otherwise, there is no treatment of the viscous fluid in DEM simulations.

\begin{figure}[ht]
\begin{center}
\centerline{\includegraphics[width=.8\textwidth]{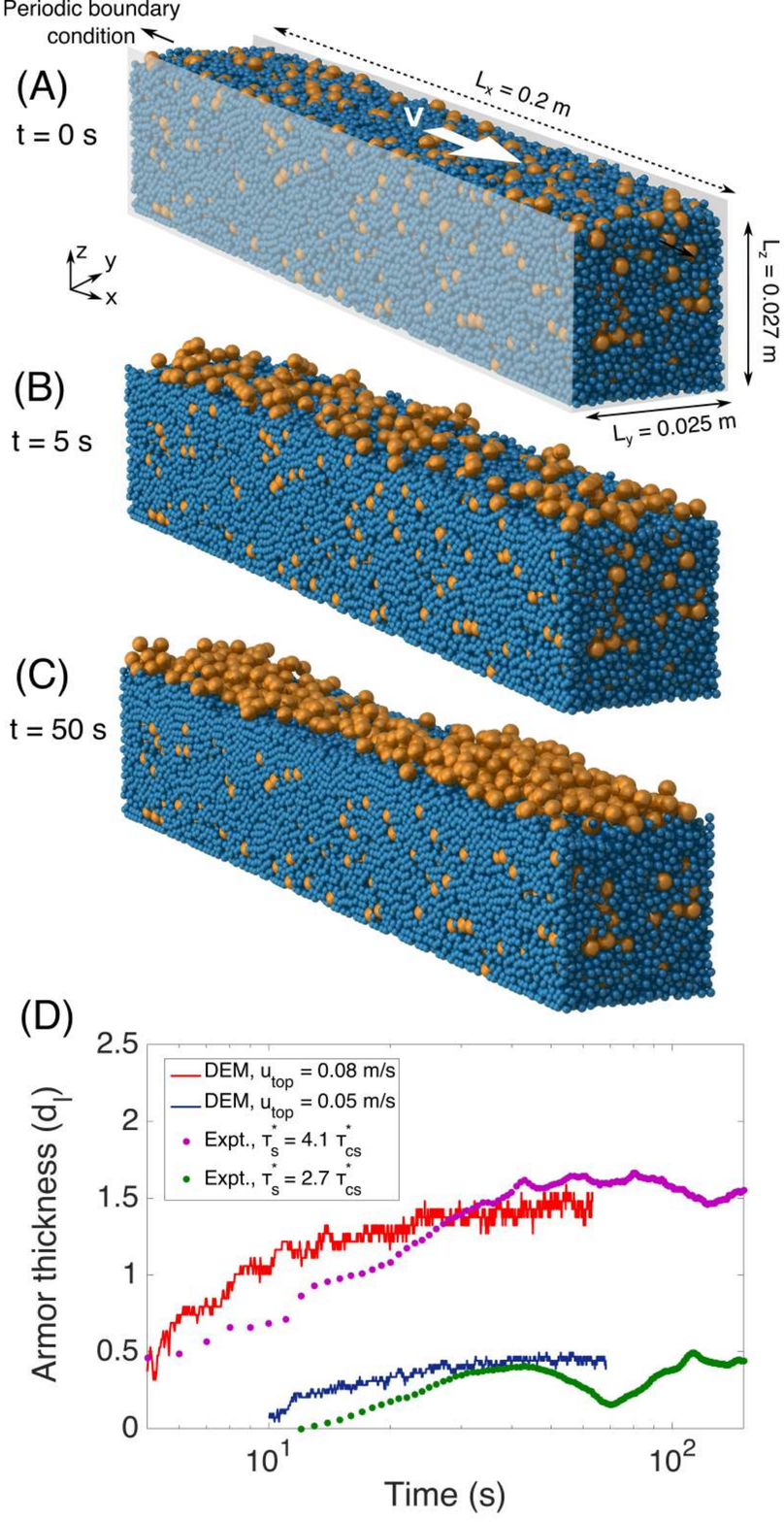}}
\caption{DEM simulation of a dry, sheared granular bed with $u_{top} = 0.08$ $m/s$ equivalent to the fluid-driven sheared granular bed at $\tau_s^* = 4.1\tau^*_{cs}$. The top layer of large grains that drives particles underneath is shown in Fig S9. (A) Model domain and initial conditions. Granular pack shown after (B) 5 $s$ and (C) 50 $s$ of shearing. Note rapid segregation, and depletion of the near-surface zone of large grains, as a consequence of bed load. (D) Evolution of armor thickness for the DEM model and experiments at two equivalent driving stresses indicated in the legend.}
\label{fig4}
\end{center}
\end{figure}

The model domain is constructed to have a geometry, grain size and size-volume ratio that are the same as the experimental setup (Fig. \ref{fig4}A). The system is driven by a layer of large grains deposited at the surface and moving at constant velocity $u_{top}$ in the $x$ direction. Simulations are run for a duration that is equivalent to $\sim 60\, \text{s}$ due to their computational expense; however, we show below that this is sufficient time to observe the fundamental dynamics. Simulations show behavior that is qualitatively comparable to the fluid-driven experiments of armoring, confirming the existence of two stages of segregation (Fig. \ref{fig4}). First is fast segregation within the rapid granular flow regime (first few grain diameters from the surface). Then, once grains are depleted from this ``bed load'' zone (Fig. \ref{fig4}C), armoring transitions to a slow stage driven by creep from deeper layers. 

For a more direct comparison, we examine the growth of armor thickness through time (see Methods) for the previous simulation and an additional run with $u_{top}=0.05$ $\text{(m/s)}$ which corresponds to $\tau_s^* = 2.7\tau^*_{cs}$. For both runs the agreement of DEM simulations with the experiments is reasonably good (Fig. \ref{fig4}D). This agreement is especially encouraging given that: simulations neglect fluid flow entirely; the initial concentration of large grains in the experiments was difficult to control and not uniform; and there was no tuning or calibration done for the DEM runs, beyond adjusting the velocity of surface grains to match experiments. 

\section*{Discussion}
\noindent
Even though our flows were laminar, experiments and theory have shown that laminar bed load is similar to its turbulent counterpart in many respects \citep{charru2004erosion,houssais2012bedload,ouriemi2009sediment, maurin2016dense,du2003granular}. Our results show armoring dynamics that are qualitatively similar to previous experiments \citep{hassan2006experiments, esurf-4-461-2016} conducted under conditions more representative of gravel rivers --- i.e., poly-disperse and natural-shaped particles (average grain diameter $d\sim 1 \; \text{cm}$) in turbulent flows with driving stress $\tau^*\approx2 \tau^{*}_{c}$. Those studies \citep{hassan2006experiments, esurf-4-461-2016} found a Shields-stress dependent armoring rate with a relatively rapid initial stage (a few hours) followed by slower stage. While data on particle motions were not reported, we can perform a scale analysis of the expected bed-load armoring timescale due to granular segregation, $t_{adv} \sim \frac{h_{bl}} {aU_{sf}}$, by assuming: $h_{bl} = (3-5)d$  \citep{devries2002bedload}; $U_{sf} \sim 1\; \text{cm/s}$ \citep{lajeunesse2010bed,drake1988bedload}; and our experimentally-determined value $a \sim 10^{-3}$. This analysis yields $t_{adv}\sim [1-2]\; \text{hrs}$, within the observed range of experiments \citep{hassan2006experiments,esurf-4-461-2016}, and may be a reasonable bed-load armoring timescale for natural gravel rivers. Translation to the field, however, may need to account for the presence of bed and bar forms that can influence armor formation \citep{marion1997experimental}.

Authors of previous experiments \citep{dietrich1989sediment, hassan2006experiments, esurf-4-461-2016} attributed armor development to a lack of sediment supply to the channel, which they hypothesize resulted in winnowing of fines and concentration of coarse grains --- in other words, sediment-supply imbalance. Our experiments, however, showed no significant size-selective transport at the surface and, more importantly, there were no supply limitations because the flume is annular. We can thus rule out sediment-supply imbalance for our experiments. Our results support the kinetic sieving model, on which the phenomenological Gray-Thornton equation is based. An important new finding, however, is that segregation does not occur only in the ``active layer''. If the bed-load zone corresponds to the active layer, then the associated sorting is important but occurs rapidly. Creep delivers grains from far below the bed-load zone to the surface, contributing to persistent armor development that was not previously recognized. The agreement of DEM simulations and experiments confirm the contention of Frey and Church \citep{frey2009river} that river-bed armoring is a granular segregation phenomenon, and suggest minimal influence of the fluid beyond determining the surface grain velocity. We point out that sediment-supply imbalance may still be important for armoring in some rivers; in particular, under sediment limitations such as downstream of dams where river beds experience net erosion that may preferentially remove finer grains \citep{dietrich1989sediment}. The granular segregation dynamics revealed here, however, would operate in all environments regardless of sediment supply and may therefore be more prevalent. Future experiments with more 'natural' flow and particle conditions, that control for sediment supply while also examining precise granular motion from the surface to the bottom of the granular pack, would be helpful for assessing the relative importance of these different mechanisms. A potential field confirmation of the armoring mechanism proposed here would be the observation of a zone underneath the armor layer that is almost entirely depleted of large grains (Figs.~\ref{fig2}C and D).  Size-selective surficial transport would not be expected to influence the concentration of coarse grains beneath the armor layer.

Our work sheds new light on the mechanics of granular segregation. Experiments clearly show that vertical advection of large grains is shear-rate dependent. Explicit accounting of this dependence, and also of shear-rate dependent diffusion, is needed in order to explain observed segregation rates for the rapid granular flow regime (i.e., bed load). Moreover, data and models demonstrate that creep contributes to segregation, and that its mechanism is distinct from rapid granular flows. Large grains in the creep zone show no preference for upward- or downward-directed motion. Their long-time motion may be modeled as vertically-isotropic and constant diffusion. Short-time dynamics show that creeping grains are caged, and indicate that their motion is likely induced by long-range transmission of forces through the granular contact network \citep{lherminier2014revealing,aste2002stress}. This may be why creep motion is independent of shear rate, at least for the range of Shields stresses examined here. It is intriguing that isotropic diffusion in creep can give rise to a net upward flux of large grains. Based on our results, we hypothesize that this flux arises because large grains that cross the boundary into bed load are then advected to the surface. If correct, this implies that a purely creeping granular pack (no bed load) should not produce armoring. 

The experimental and modeling results presented here are a first step in assessing the contribution of fast and slow particle motion to vertical segregation. Our sediment mixture was bi-disperse in order to establish connections between granular shear segregation and river-bed armoring, but many systems of interest (including rivers) have a polydisperse grain size distribution that may exhibit different behavior. Such a distribution would challenge the application of continuum models, but is amenable to experimental and DEM simulation approaches. River-bed armoring in our experiments and models was found to be driven by bottom-up granular segregation, rather than top-down surficial sorting driven by the fluid. Our findings show how information from the surface, in terms of fluid-driven shear, is transmitted deep into the subsurface through grain-grain interactions that are typically neglected in sediment transport models. Granular motion in the subsurface transmits information back to the surface through the delivery of coarse grains, linking surface dynamics to subsurface structure. By examining the river bed as a discrete medium, we were able to link the macroscopic pattern of armor development to the physics of sheared granular systems. Our results add to a growing body of evidence that sediment transport systems belong to a broader class of granular flows \citep{frey2009river,capart2011transport,houssais2015onset,houssais2015rheology}, and show how examining geophysical flows through the lens of granular physics can reveal novel insights for both fields. 

\section*{Materials and Methods}
\noindent
Details of the experimental setup and shearing protocol are described in SI section 1. The bed preparation protocol was inspired by Golick and Daniels \citep{golick2009mixing}. Grains were initially deposited in an inverse-segregated state, with large grains at the bottom, and then subject to a driving stress equivalent to $\tau_s^* = 20\tau^*_{cs}$ for $\sim$ 1 minute to fully suspend and mix the large and small populations. Fluid shear was halted and the suspension left for $\sim$ 30 minutes to allow sedimentation, relaxation and compaction of the granular bed to reach completion (Fig. S1). 

Implementation of the continuum model is described in SI section 6 (Figs. S3-S8; Table S1). Details of the DEM model and parameters appear in SI section 7 (see Table S2). The local concentration of large grains is defined as, $\phi_l(z) = \frac{\langle A_l \rangle_x}{\langle A_l +  A_s \rangle_x}$ where $A_l$ and $A_s$ are large and small grains areas, respectively, and  $\langle \cdot \rangle_x$ indicates pixel-wise streamwise integration.\\}

\section*{Acknowledgments}
Research was supported by US Army Research Office–Division of Earth Materials and Processes grant 64455EV, US National Science Foundation (NSF) grant EAR-1224943, NSF INSPIRE/EAR-1344280, and NSF MRSEC/DMR-1120901. BF is a synthesis postdoctoral fellow of the National Center for Earth-surface Dynamics (NCED2 NSF EAR-1246761).

\section*{Supplementary Information}
\noindent
\setcounter{figure}{0}
\renewcommand{\thefigure}{S\arabic{figure}} 
\renewcommand{\thesubsection}{\arabic{subsection}}   
\renewcommand{\thetable}{S\arabic{table}}  
\subsection{Experimental setup and protocol}
To generate 2D images of our 3D experimental system, we index-matched PMMA particles ($d_s =1.5\, \text{mm}$, $d_l = 3.0\, \text{mm}$, Engineering Laboratories) with a mixture of viscous oils (85\% of PM550 and 15\% of PM556 from Dow Corning), and excited a dye (Exciton, pyrromethene 597) dispersed in the oil with a green laser sheet (517 nm, 50mW) of thickness $\simeq d/10$. The experiment was conducted on a vibration-damping optical table, while a damping coupler was used to connect the driving motor to the flume.  

\begin{figure*}
  \centering
    \includegraphics[width=1.0\textwidth]{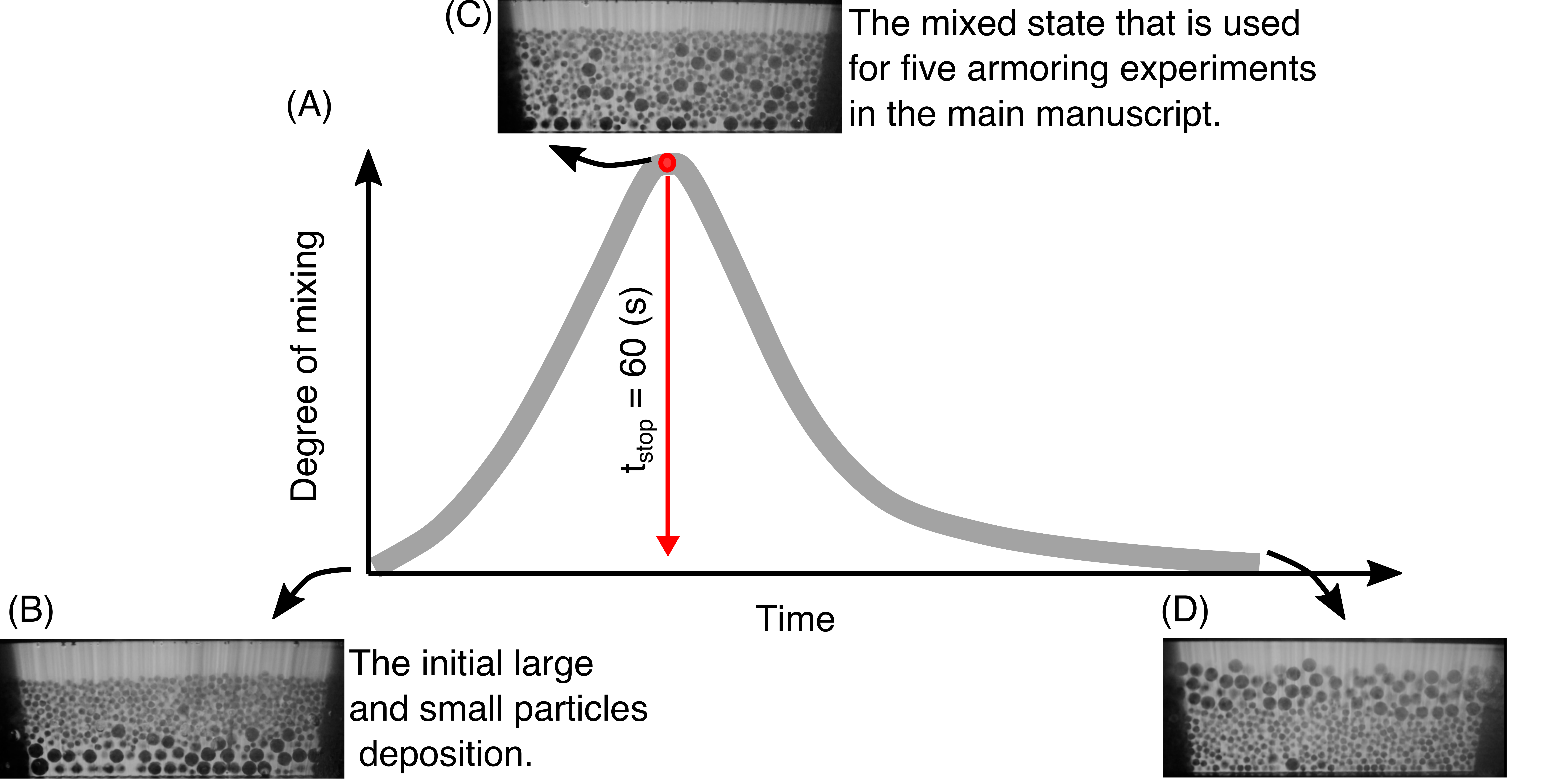}
   \caption{(A) A schematic for the preparation protocol used for preparing the initial bidisperse mixture of particles. Small and large particles are initially deposited in an inversely segregated manner (panel B). The initial deposition has been performed by gently pouring first large particles and then small particles uniformly but manually from a very close distance to the bottom of the chamber. The system is then subjected to a rotation of $\Omega= 22\, \text{r.p.m.}$ that is a driving stress equivalent to $\tau_s^* = 20\tau^*_{cs}$ for $\sim$ 1 minute in order to mix the large and small populations (panel C). This shear stress was sufficient to fully suspend all particles in the channel. Fluid shear was then stopped completely and the suspension was left for $\sim$ 30 minutes to allow time for sedimentation, and relaxation and compaction of the granular bed, to reach completion. The final random packed layer at the end of preparation protocol has a thickness $\sim 15.5 d_s$ for all experiments. This state is used for the five Shields number armoring experiments reported in the main manuscript. (C) A hypothetical fully segregated state that one could obtain if continuing shearing the initial sample at the large preparation shear stress of $\tau_s^* = 20\tau^*_{cs}$ for about 3 minutes.} 
  \label{figs1}
  \end{figure*}

The granular bed for each experiment was prepared with the same protocol: grains were initially deposited in an inverse-segregated state, with large grains at the bottom, and then subject to rotation $\Omega= 22\, \text{r.p.m.}$ that is a driving stress equivalent to $\tau_s^* = 20\tau^*_{cs}$ for $\sim$ 1 minute in order to mix the large and small populations. This shear stress was sufficient to fully suspend all particles in the channel. Fluid shear was then stopped completely and the suspension was left for $\sim$ 30 minutes to allow time for sedimentation, and relaxation and compaction of the granular bed, to reach completion. Figure~\ref{figs1} provides further information about the preparation protocol. The final random packed layer at the end of the preparation protocol had a thickness $\sim 15.5 d_s$ for all experiments. Then, a constant rotation $\Omega$ drove the system during the entire experiment. The duration of the experiment was not fixed; each lasted long enough ($24\, \text{hr}$ or longer) for all particles present in the recorded frames to exhibit some significant displacement during the run. We computed the fluid-flow depth $h_f = H_f-z_s$, where $H_f$ is the total depth of the
flume and $z_s$ is the elevation of the surface as described below. We compute the fluid-flow velocity at the top plate in the channel center as
$U_f=\Omega.2\pi R$, where $R = 17\, \text{cm}$ is the radial distance to the channel center. The fluid boundary-shear stress is then calculated as $\tau =  \eta U_f/h_f$.  For our definition of $\tau^*$, and throughout our analysis, we assume the fluid flow is laminar and unidirectional in the azimuthal direction of the annular flume. The laminar assumption is justified because the Reynolds number associated with the fluid channel above the bed is small. We estimate this Reynolds number as $Re = \frac{\rho U_{plate} h_f}{\eta}$, which is $\approx 4$ for the largest $\Omega$ in these experiments. The unidirectional assumption is justified based on the small ratio of radial viscous stress to the azimuthal viscous stress for our experimental conditions:

\begin{equation} \label{eq:sgs1}
\frac{\text{Radial stress}}{\text{Azimuthal stress}} = c Re \frac{h_f}{R} = 0.4\%
\end{equation}

where $h_f\simeq3d_s$, $R$ is the flume radius and $c \simeq 0.06$ is an estimated coefficient \citep{charru2004erosion} that is only weakly dependent on the flow aspect ratio.

\subsection{Detection of the bed surface} 
\noindent
In order to detect the surface position, first the concentration profile $C(z)$ for a given configuration of particles is determined from a processed binary image, valued at zero outside of particles and one inside of particles. For each elevation $z$, the concentration is determined as the pixel-wise average in the $x$ direction. This concentration is the $1D$ analogue of packing fraction, the fraction of space occupied by the particles. The surface is defined as the position $z_s$ at which the concentration crosses fifty percent of its saturated value \citep{duran2012numerical, houssais2015onset}. We use a fixed threshold of 0.35 to define the surface position, as the saturated value does not vary significantly from experiment to experiment. We define $z_s$ after averaging the concentration for a $\Delta t =100\, \text{s}$ at the beginning of each experiment. The time duration is sufficiently long for the flux convergence time as observed in our earlier study \citep{houssais2015onset}. Slow granular compaction \citep{ben1998slow,richard2005slow} and slow dilation due to segregation [30] approximately counterbalance such that the surface position remains almost constant as the armoring experiments progress. The bed surface position is used for calculating the Shields stress at each experiment.

\subsection{Imaging technique and particle detection/tracking} 
\begin{figure*}
  \centering
    \includegraphics[width=0.4\textwidth]{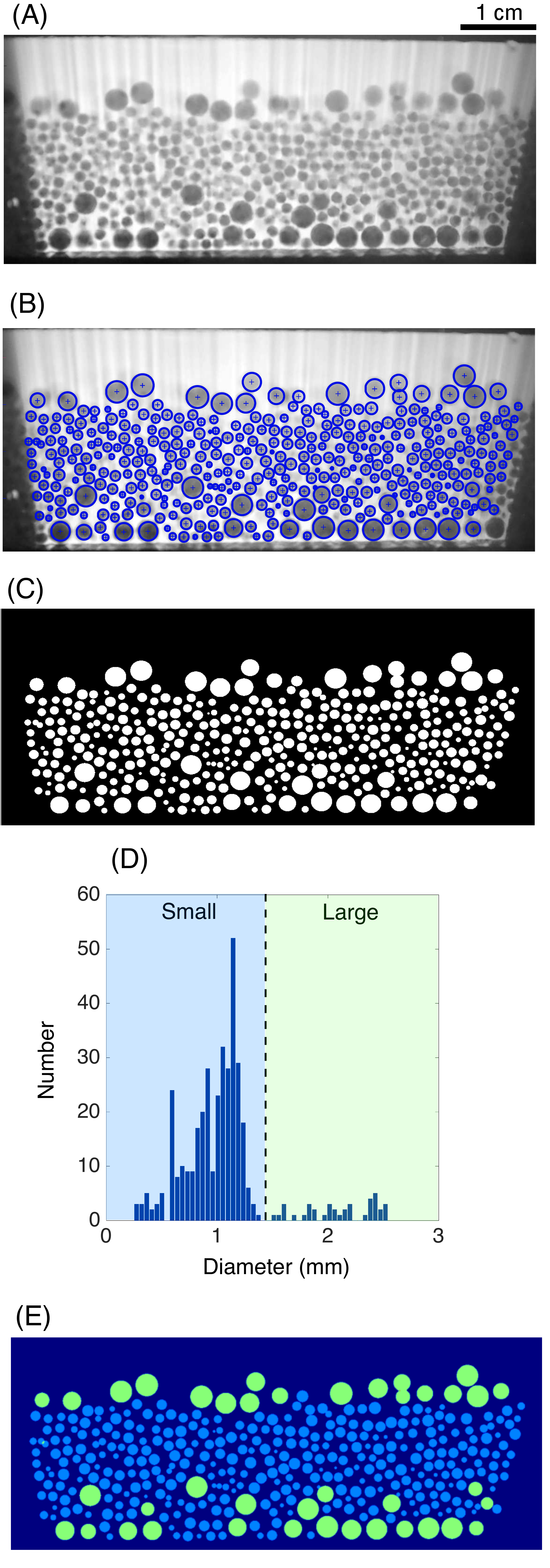}
    \caption{(A) An example raw image from the experimental run. (B) Detected particles for panel (A) after processing of the image and running our particle detection algorithm. (C) Binary image of particles detected in panel (B). Time sequences of similar images are used for calculating streamwise velocity profiles using cross-correlation analysis. (D) Size distribution of detected particles, diameter threshold for small and large particles and the resulting subsets. (E) Detected small and large particles for the example snapshot in panel (A). This is the final result of the image analysis and particle detection, and similar images are used for all post-processing and further analysis presented in the main manuscript.}
  \label{figs2}
  \end{figure*}

\noindent
Using a Nikon DSLR 5100 digital camera, we record the real-time positions of single particles by acquiring the fluorescence intensity from a laser dye (concentration $\approx 1 \mu M$) dispersed in the fluid that is suitable for long data acquisition without significant photobleaching. To sample fast dynamics near the surface, where the relevant timescale is the settling time of particles over their own diameter $d/v_{sed} = 0.68 \text{s}$, we acquire images continuously at $24 \;\text{Hz}$ for $10-20\, \text{mins}$. To sample slow dynamics in the bed, we acquire single images at a rate of one every $15 \text{s}$ for $24 \text{hr}$ or longer. Figure~\ref{figs2}A shows a sample raw image at the start of an experiment. To detect the positions of the particles with subpixel accuracy, we find particle positions to pixel accuracy by peak-finding above a threshold. The details of the background correction process are described in the supplementary materials of our previous publication \citep{houssais2015onset}. The process is designed specifically to handle both long-wavelength background intensity variations and intermediate wavelength fluctuations (stripes) due to slight mismatches in the index of refraction of the particles relative to the fluid. After removing the background, we determine sub-pixel positions using the radial symmetry method \citep{loy2003fast}. We use this method to identify radially symmetric features at different probing length scales. Then, features are identified as peaks in the three-dimensional image $(x,y,R)$, over a range of radii that is sampled linearly in log-radius. Peaks are obtained by two passes, first a pixel-scale search, then a local quadratic fit to obtain subpixel positions. A snapshot of detected particles with this method is superimposed on a gray-scale raw image and is shown in Figure~\ref{figs2}B. The same detected particles are also shown in binary format in Figure~\ref{figs2}C. A fixed diameter threshold of $d=1.38 \text{mm}$ is used for separating large and small particles in all experiments as shown in Figure~\ref{figs2}D. Finally, a snapshot of identified large and small particles using this threshold is shown in Figure~\ref{figs2}E.

We stitch positions at different frames into tracks using Lagrangian particle tracking \citep{ouellette2006quantitative}. 

\subsection{Velocity profiles}
\noindent
For each experiment, a 10 minute video capture with frame rate 24 fps at the start of the experiments is converted and processed into consecutive binary images following the procedure described in imaging section above. The consecutive binary images, $I(t)$ and $I(t+\Delta t)$ are then used as the input of pixel-wise cross-correlation analysis along the $x$ direction at each pixel elevation $z$. The position of the central peak in the cross-correlation between $I(t)$ and $I(t+\Delta t)$ corresponds to the average streamwise distance traveled by gains at elevation $z$ during $\Delta t$ without regard to small and large particle species, {\it i.e.} for all particles. The result is averaged over the full duration of the video capture. This technique yields a time-averaged streamwise velocity profile $u_x(z)$ for all particles. Note, large particles are weighted more heavily than small particles. The results are in agreement with velocity profiles determined from the particle-tracking method for all experiments. For the case of Figure 2A, the velocity profile of large grains is computed using the particle tracking method described in the imaging technique section above. 

\subsection{Determination of armor thickness}
\noindent
The top surface of the armor layer, $z_{sa}$, is characterized as the position where the streamwise averaged concentration of large grains equals $\phi_l = 0.9$. The interface (bottom) of the armor layer with the rest of the granular bed, $z_{i}$, is calculated as the location where the gradient of $\phi_l$ reaches a minimum below the surface. The surface and interface positions time-series are smoothed using a running average of temporal window size $8.33\, \text{s}$ for images obtained from video capture conversions and temporal window size $833\, \text{s}$ for image captures.
These are shown with dashed lines in Fig.~2C. The thickness of the armored layer is defined as  $z_{sa} - z_{i}$.

\subsection{Implementation of continuum model}
\begin{figure*}
  \centering
    \includegraphics[width=0.4\textwidth]{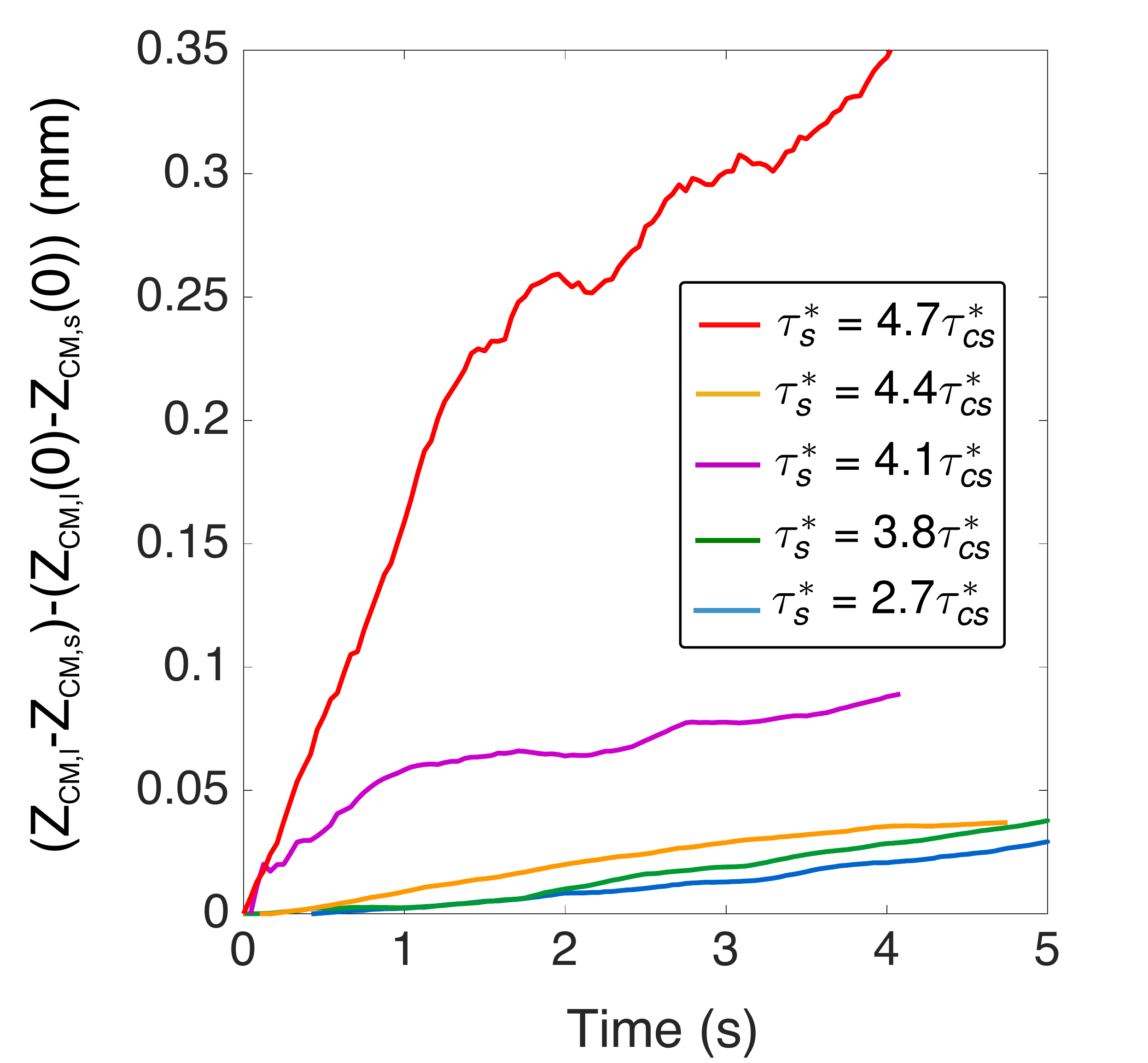}
    \caption{Relative position of the vertical ($z$) component of the center of mass of the assembly of large and small particles at the start of the experimental runs for five Shields numbers. Here, $Z_{CM,l}$ and $Z_{CM,s}$ are the $z$ component of the center of mass position of the assembly of large and small particles. $Z_{CM,l}(0)$ and $Z_{CM,s}(0)$ denote the initial positions of the center of masses of two assemblies, i.e. the positions at $t=0$.} 
  \label{figs3}
  \end{figure*}

\begin{figure*}
  \centering
  \includegraphics[width=1.0\textwidth]{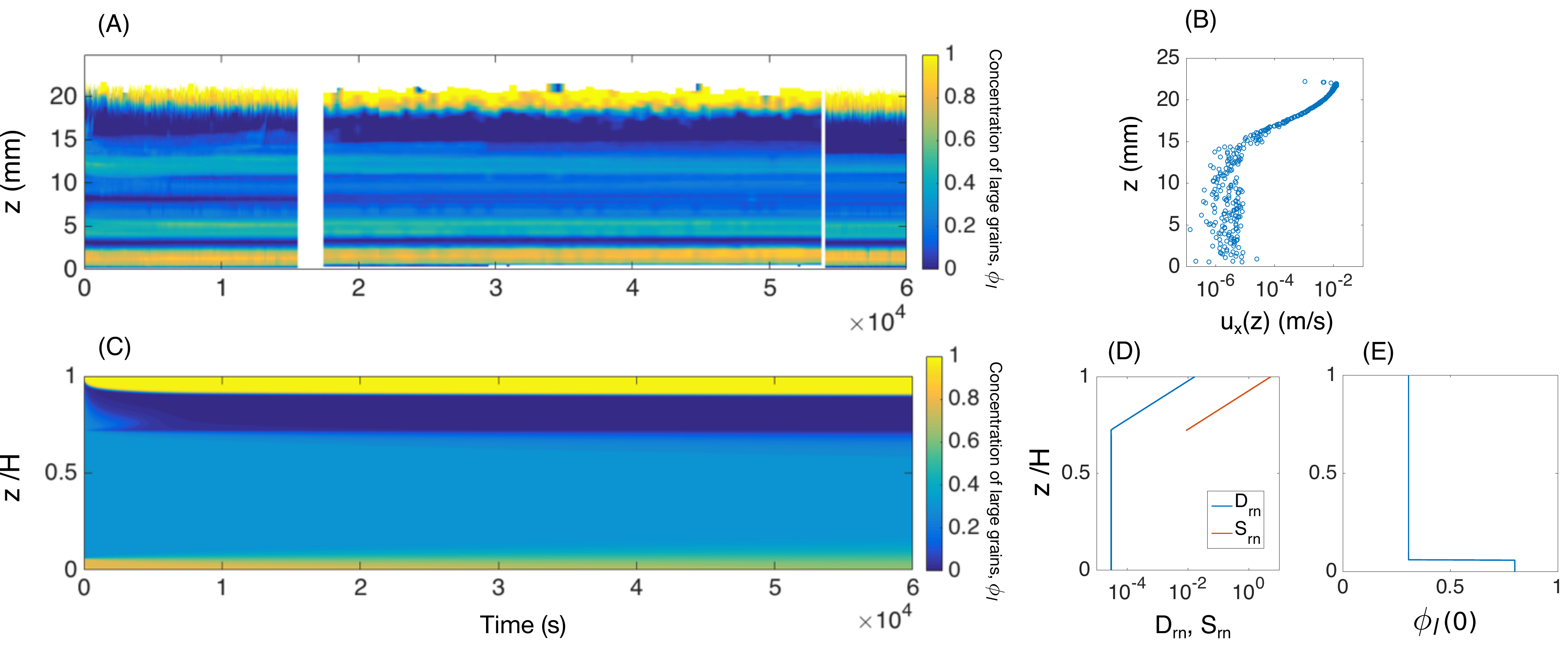}
  \caption{(A) 1D ($x$-averaged) concentration map of large grains over time for shear stress $\tau_s^* = 2.7\tau^*_{cs}$. (B) Streamwise velocity profile ($u_x(z)$) of the granular bed for shear stress $\tau_s^* = 2.7\tau^*_{cs}$. (C) 1D ($x$-averaged) concentration map of large grains over time from the advection-diffusion model, with velocity profiles and initial condition corresponding to the shear stress $\tau_s^* = 2.7\tau^*_{cs}$ in panel (A). (D) Vertical profiles of $S_{rn}$ and $D_{rn}$ that were implemented in the continuum model. (E) Initial vertical profile of concentration of large grains ($\phi_l(t=0)$)}
  \label{figs4}
  \end{figure*}

\begin{figure*}
    \centering
    \includegraphics[width=1.0\textwidth]{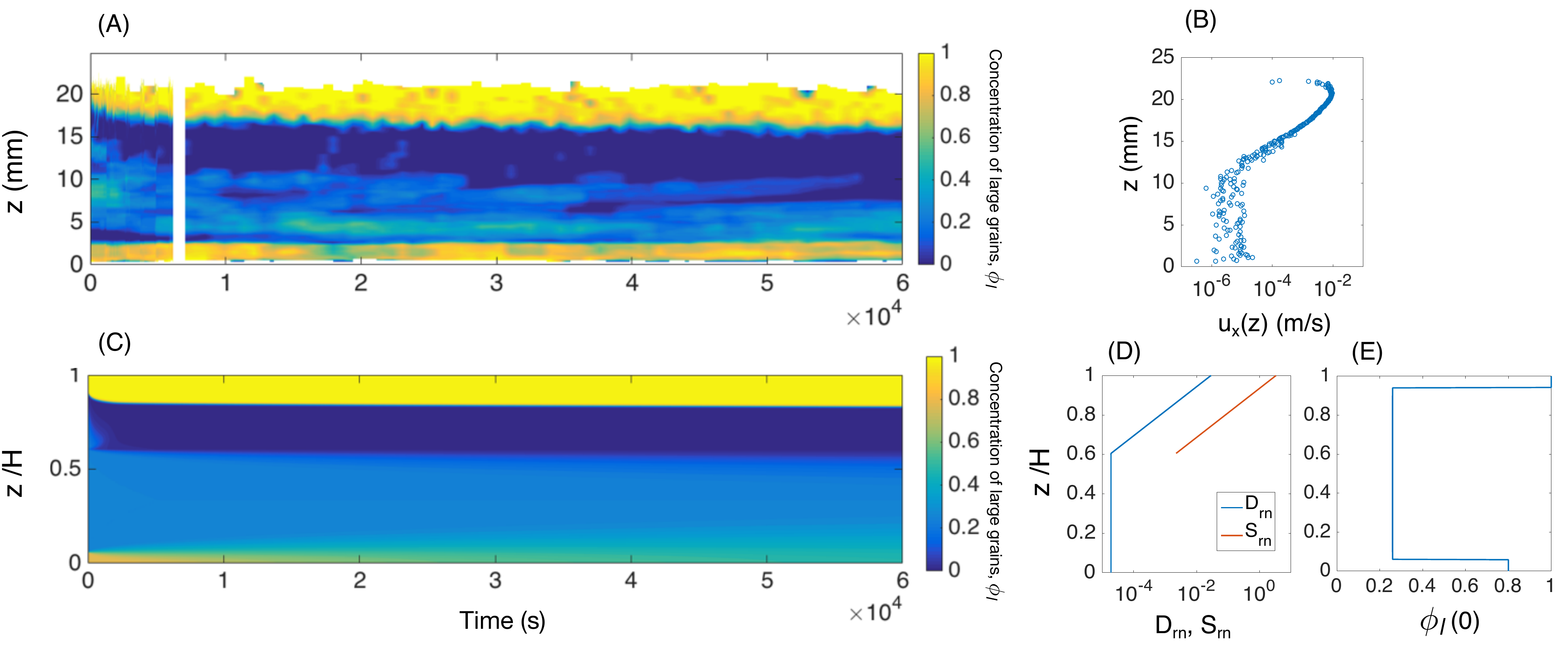}
      \caption{Concentration map for large grains for $\tau_s^* = 3.8\tau^*_{cs}$ experiment. All panels follow Figure S4.} %
      
  \label{figs5}
  \end{figure*}

\begin{figure*}
  \centering
  \includegraphics[width=1.0\textwidth]{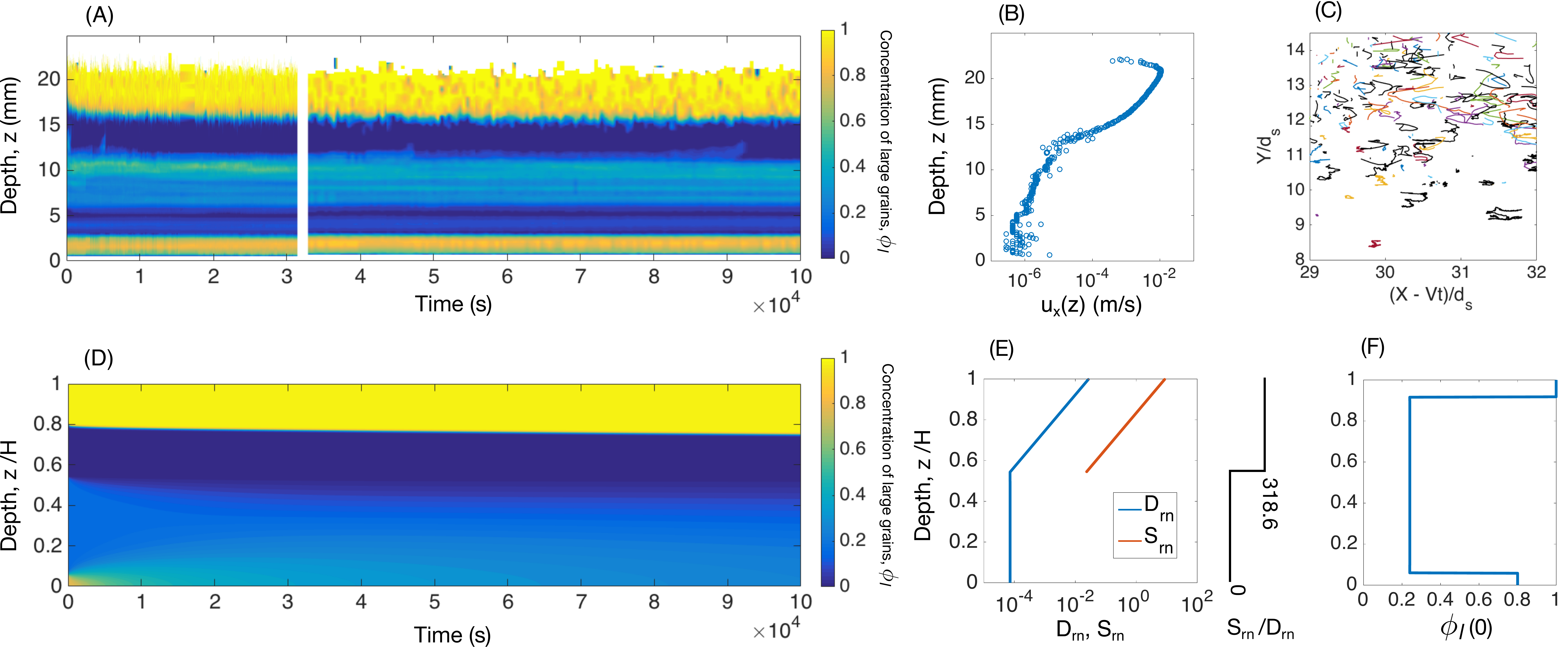}
  \caption{Concentration map for large grains for $\tau_s^* = 4.1\tau^*_{cs}$ experiment. All panels follow Figure S4.}
  \label{figs6}
  \end{figure*}

\begin{figure*}
    \centering
    \includegraphics[width=1.0\textwidth]{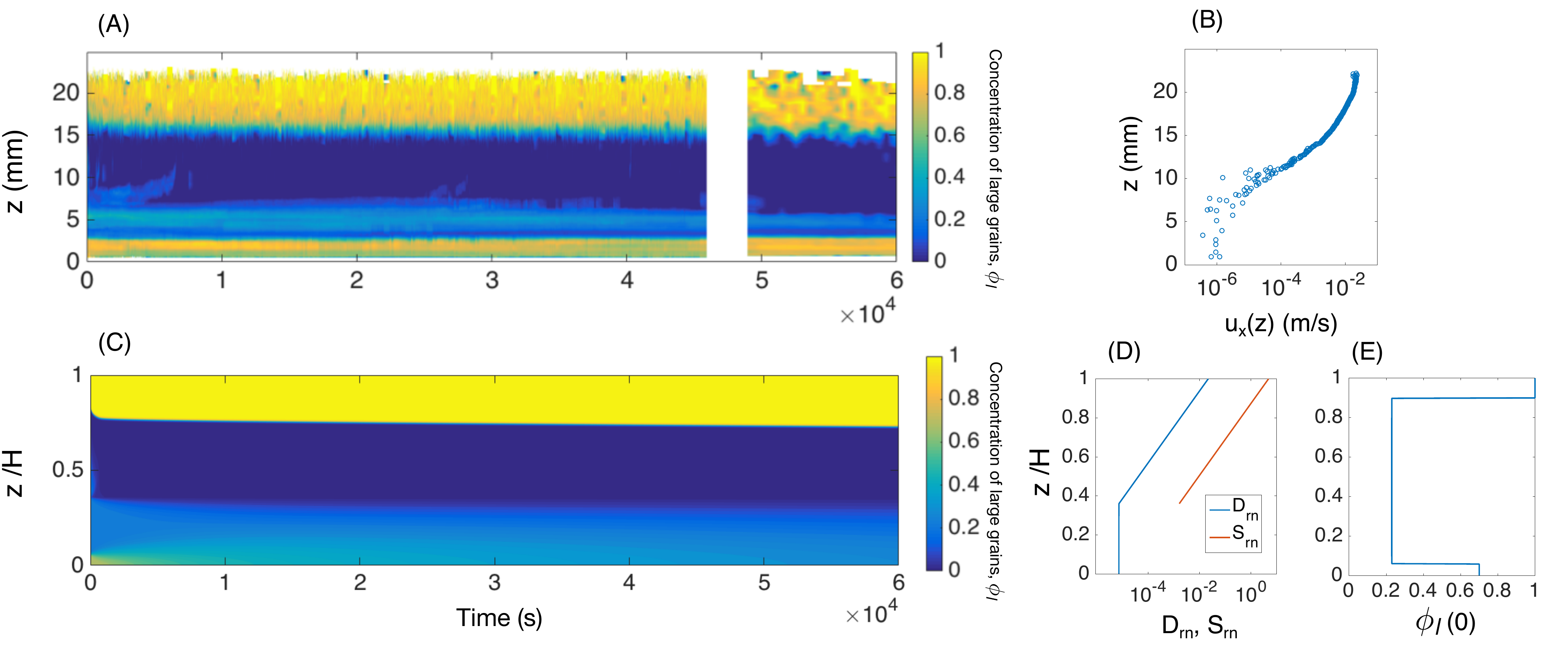}
      \caption{Concentration map for large grains for $\tau_s^* = 4.4\tau^*_{cs}$ experiment. All panels follow Figure S4.}
  \label{figs7}
  \end{figure*}

\begin{figure*}
    \centering
    \includegraphics[width=1.0\textwidth]{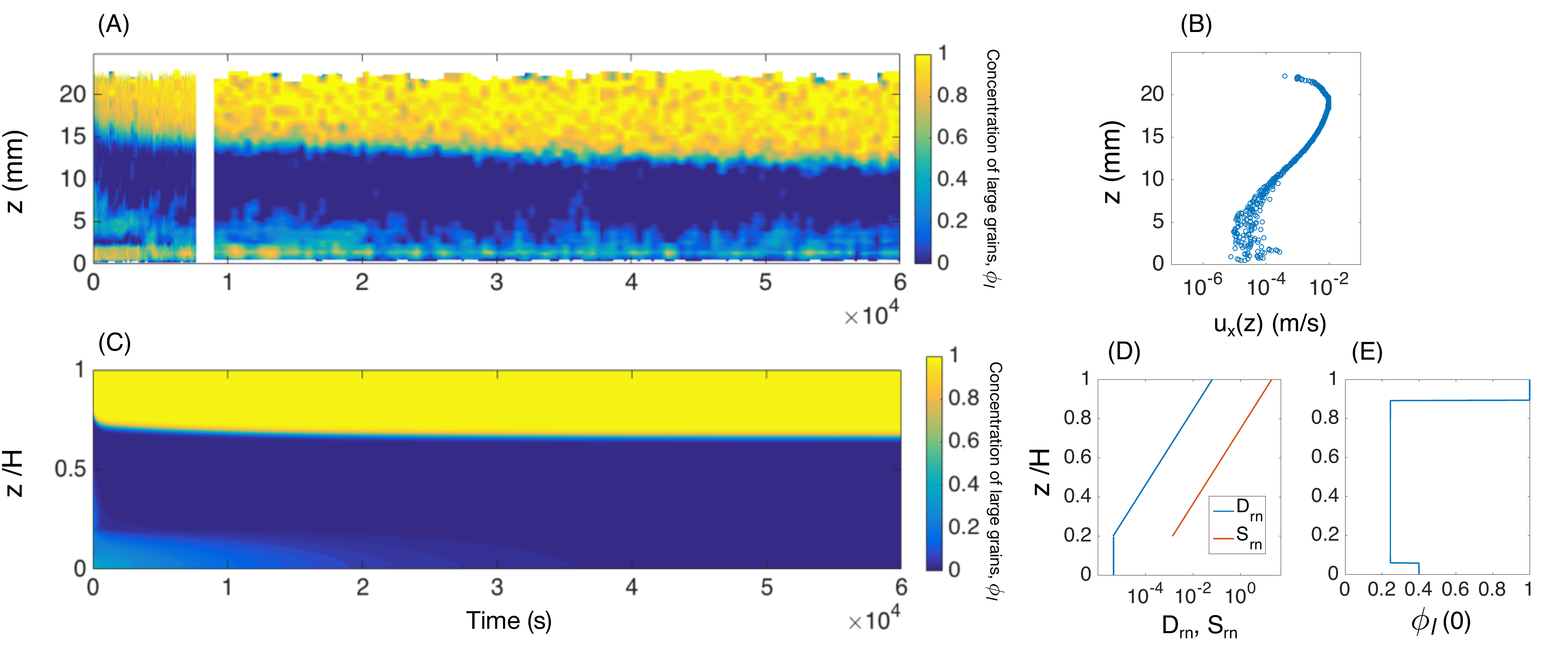}
      \caption{Concentration map for large grains for $\tau_s^* = 4.7\tau^*_{cs}$ experiment. All panels follow Figure S4.}
  \label{figs8}
  \end{figure*}

\noindent
The variables used to compute the advection and diffusion parameters for each experiment are reported in Table [\href{https://www.dropbox.com/s/hemoeikexg6fbcb/main.pdf?dl=0}{S1}]. The maximum bulk segregation velocity, $q$, for each experiment is measured from relative displacement of the vertical $z$ component of the center of mass position of large particles $Z_{CM,l}$ relative to small particles $Z_{CM,s}$. The data for the relative $Z_{CM}$ displacement for all five stresses is presented in Figure~\ref{figs3}. The initial concentration profile of large particles $\phi_{l}(0)$ is determined from the first $10$ $\text{s}$ of each experimental run; a simplified version of it is used as another input to the PDE model (Figs.~\ref{figs4} to \ref{figs8}). The time-averaged streamwise velocity profiles for each experiment are reported in Figs.~\ref{figs4} to \ref{figs8} and are used to estimate the value of $\beta$. We use a numerical implementation of the method of lines solution to solve the PDE equations and use $N_{ts}=10000$ time steps for the full experimental time ($\sim 10^{5}\text{s}$). Comparisons of the concentration maps for the PDE model and experiments, for the four additional driving stresses, are presented in the supplementary materials (Figs.~\ref{figs4} to \ref{figs8}).

\begin{table}[ht]
\centering
\caption{Parameters of the PDE simulations for five Shields numbers studied in the main manuscript.}
\begin{tabular}{l l l r}
Shields number, $\tau_s^*$            & $\beta$ value & $U_{sf}\, \text{(m/s)}$  & $\langle v_x \rangle\, \text{(m/s)}$  \\
\hline
$ 2.7\tau^*_{cs}$  & 2.30e1  & 7.0e-3 & 1.1e-3 \\
$ 3.8\tau^*_{cs}$  & 1.85e1  & 1.5e-2 & 1.4e-3 \\
$ 4.1\tau^*_{cs}$  & 1.30e1  & 6.0e-2 & 3.3e-3 \\
$ 4.4\tau^*_{cs}$  & 1.25e1  & 8.5e-2 & 5.2e-3 \\
$ 4.7\tau^*_{cs}$  & 1.20e1  & 9.0e-2 & 6.2e-3 \\

\label{tab1}
\end{tabular}
\end{table}

\subsection{Implementation of DEM}
The DEM model consists of a shear cell with sizes $0.027 \times 0.025$ $\text{m}$ in the $y \times z$ directions, and has a length $0.2 \text{m}$ in $x$ direction where periodic boundary conditions are applied. The lateral sides in the $x-z$ plane and the lower boundary in the $x-y$ plane are smooth and frictional walls, with the same mechanical and frictional properties as the grains (Table S2).

\begin{table}[ht]
\centering
\caption{Simulation parameters}
\begin{tabular}{l r}
Parameter              & Value \\
\hline
Grain density, $\rho$ & 1190 $\text{kg}/\text{m}^\text{3}$  \\
Grain diameters, d            & [$d_s$ = 0.0015, $d_l$ = 0.003] $\textrm{m}$  \\
Gravitational acceleration, $\vec{g}$  & 9.81 $\text{m}/\text{s}^\text{2}$  \\
Young's modulus, $E$     & $5\times 10^{6}$ $\text{N}/\text{m}^\text{2}$  \\
Poisson ratio, $\nu$    & 0.45  \\
Friction coefficient, $\mu$    & 0.5  \\
Coefficient of restitution, $e_n$     & 0.01  \\
Time step, $\Delta t$     & $2\times10^{-6}$ s  \\
Shear velocity, $u_{top}$ & [0.05, 0.08] $\text{m}/\text{s}$\\
\label{tab1}
\end{tabular}
\end{table}

The top side in the $x-y$ plane is open. The cell is filled with N =38812 grains that are initially inserted randomly in the cell with a desired volume fraction of 0.45. It is then equilibrated under gravitational forces for 10 million time steps equivalent to $\Delta t = 20$ s. The initial concentration of large grains is uniform in the simulation domain. The grains are free to move in other directions ({\it e.g.} to dilate) in order to resemble a free-surface and shear-driven system. The grains are modeled as compressible spheres of diameter $d_{s,l}$ that interact when in contact via the Hertz-Mindlin model \citep{johnson1987contact,landau1959theory,mindlin1949compliance}:

\begin{equation} \label{eq:sg16}
F=(k_n \delta \vec{n}_{ij}- \gamma_n \vec{v}\vec{n}_{ij})+(k_t \delta \vec{t}_{ij} - \gamma_t \vec{v}\vec{t}_{ij})
\end{equation}

where the first term is total normal force, $\vec{F}_n$, and the second term is total tangential force, $\vec{F}_t$. In Equation~\ref{eq:sg16}, $k_n$ and $k_t$ are normal and tangential stiffness respectively, $\delta$ is the overlap between grains, $\gamma_n$ is the normal damping, $v$ is the relative grain velocity, $\vec{n}_{ij}$ is the normal vector at grain contact, $\vec{t}_{ij}$ is the tangential vector at grain contact, $\gamma_t$ is the tangential damping. The full model implementation is available on the LAMMPS/LIGGGHTS webpage and several references \citep{zhang2005jamming,silbert2001granular,brilliantov1996model}. In accord with the low Stokes number of our laboratory experiments, the restitution coefficient is chosen to very small ($e_n = 0.01$) such that collisions are highly damped (Table S2). Otherwise, there is no treatment of the viscous fluid in DEM simulations. The DEM model system is frictional, meaning that the coefficient of friction, $\mu$, is the upper limit of the tangential force through the Coulomb criterion $F_t = \mu F_n$. The tangential force between two grains grows according to non-linear Hertz-Mindlin contact law until  $ F_t/ F_n = \mu$ and is then held at $F_t = \mu F_n$  until the grains lose contact. The values of density, grain diameter, Poisson's ratio and acceleration due to gravity are chosen to match the experimental conditions. The values for coefficient of restitution and friction coefficient are chosen to mimic the effects from interactions with the fluid. The Young's modulus of the particles used here is chosen to be low (Table S2), MPa rather than GPa, in order to increase the calculation time step and decrease computational cost; however, since the system is not under significant confining pressure, a softer grain-grain interaction will not have considerable effect on the results, and the simulation remains in the hard-sphere limit. The particles are sufficiently hard that we find no additional rescaling of time is necessary. The damping coefficients $\gamma_n$ and $\gamma_t$ are determined within the implementation of LIGGGHTS from the chosen value for the restitution coefficient, $e_n$. 

  \begin{figure*}
    \centering
    \includegraphics[width=0.4\textwidth]{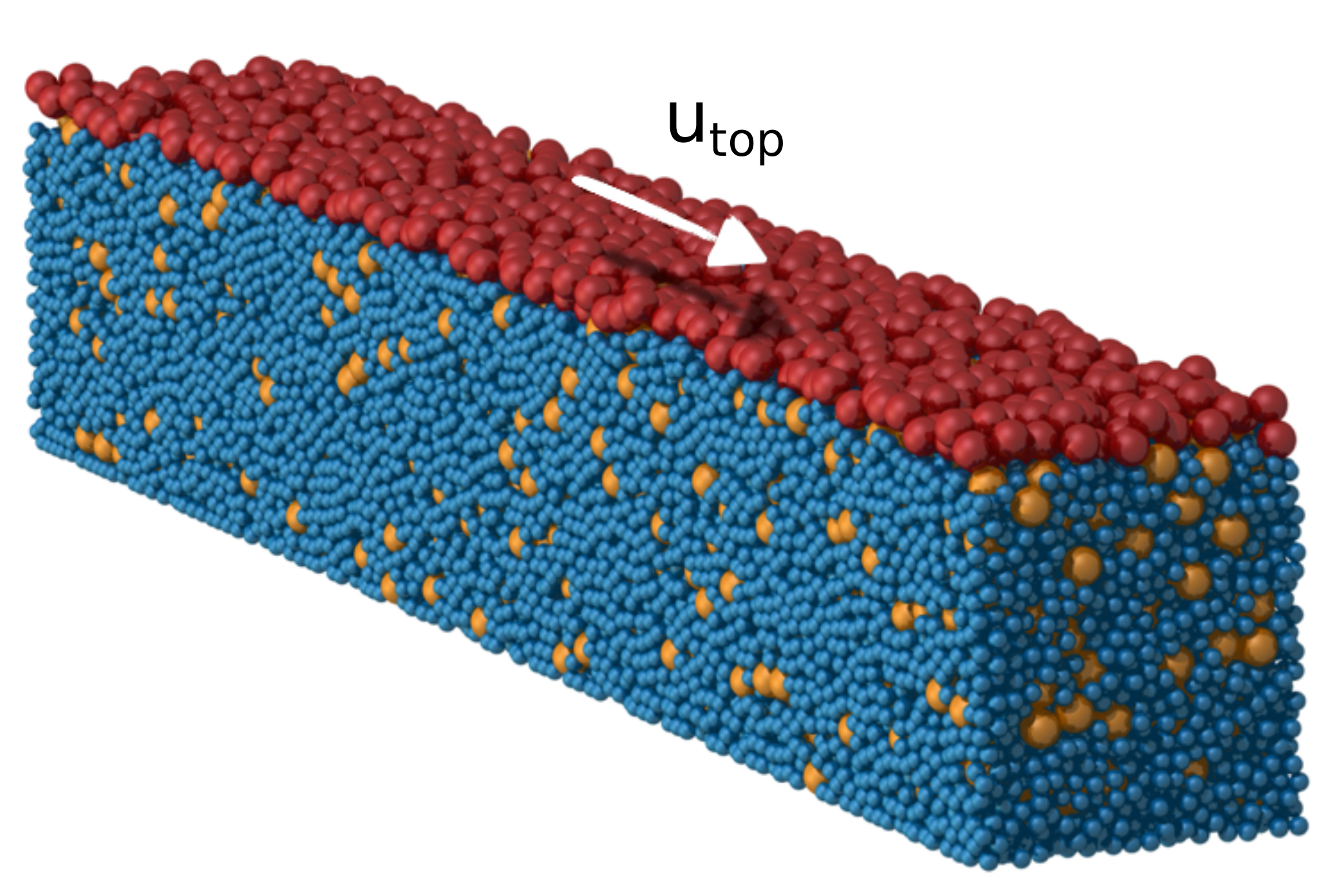}
      \caption{A snapshot from the initial conditions of the numerical DEM simulation that shows the layer of large grains deposited at the surface and moving at constant velocity $u_{top}$.}
  \label{figs9}
  \end{figure*}

\setcounter{figure}{0}

  \begin{figure*}
   \renewcommand\figurename{Movie.}
    \centering
    \includegraphics[width=1.0\textwidth]{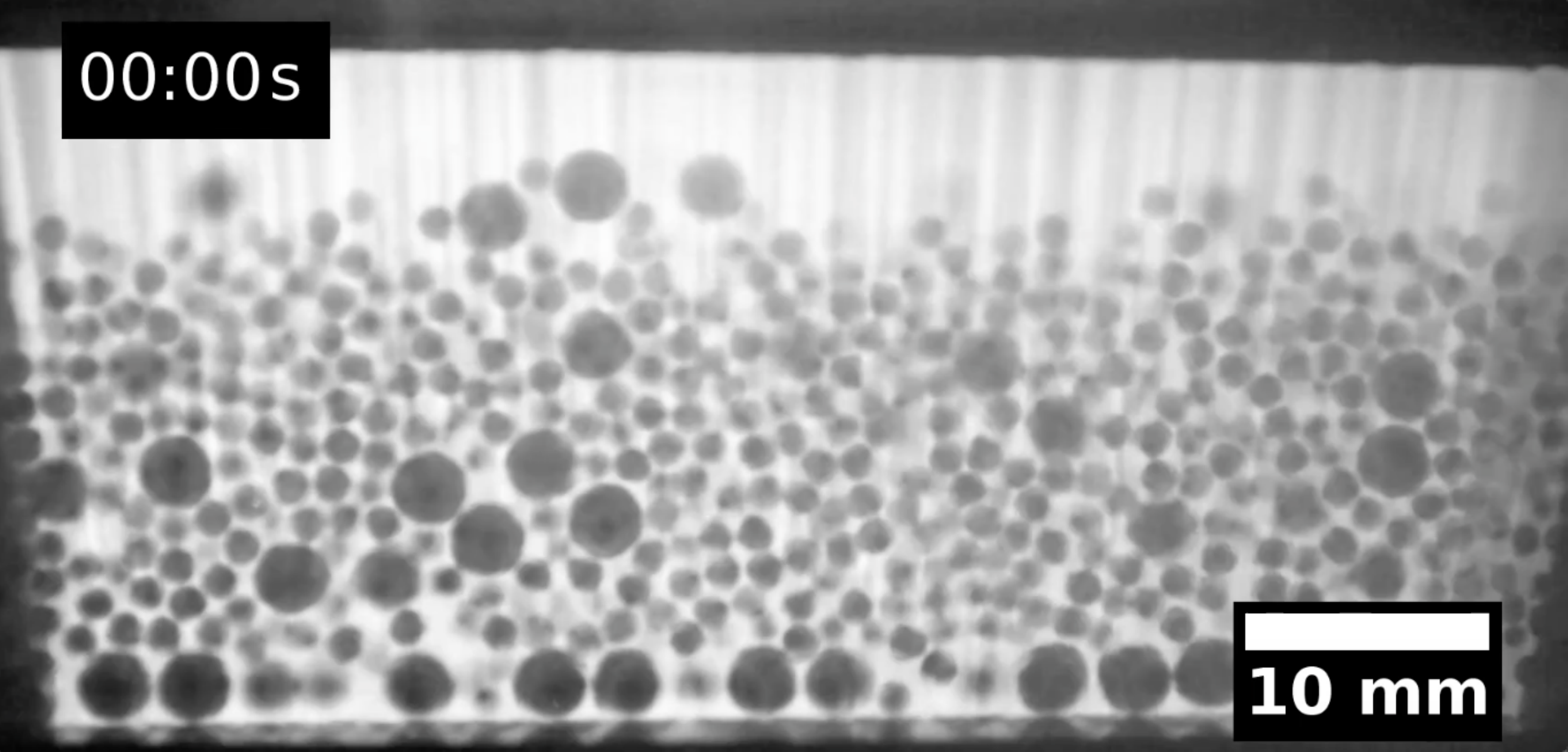}
      \caption{Real-time video of the first 30 seconds of the armoring experimental run at shear stress $\tau_s^* = 3.8\tau^*_{cs}$. The real duration of the video is 30 seconds, the same as its playback time.\\
\href{https://www.dropbox.com/s/nabr47jripqmhem/bidisperse_realtime.mp4?dl=0}{Movie S1}}
  \label{mov1}
  \end{figure*}

  \begin{figure*}
   \renewcommand\figurename{Movie.}
    \centering
    \includegraphics[width=1.0\textwidth]{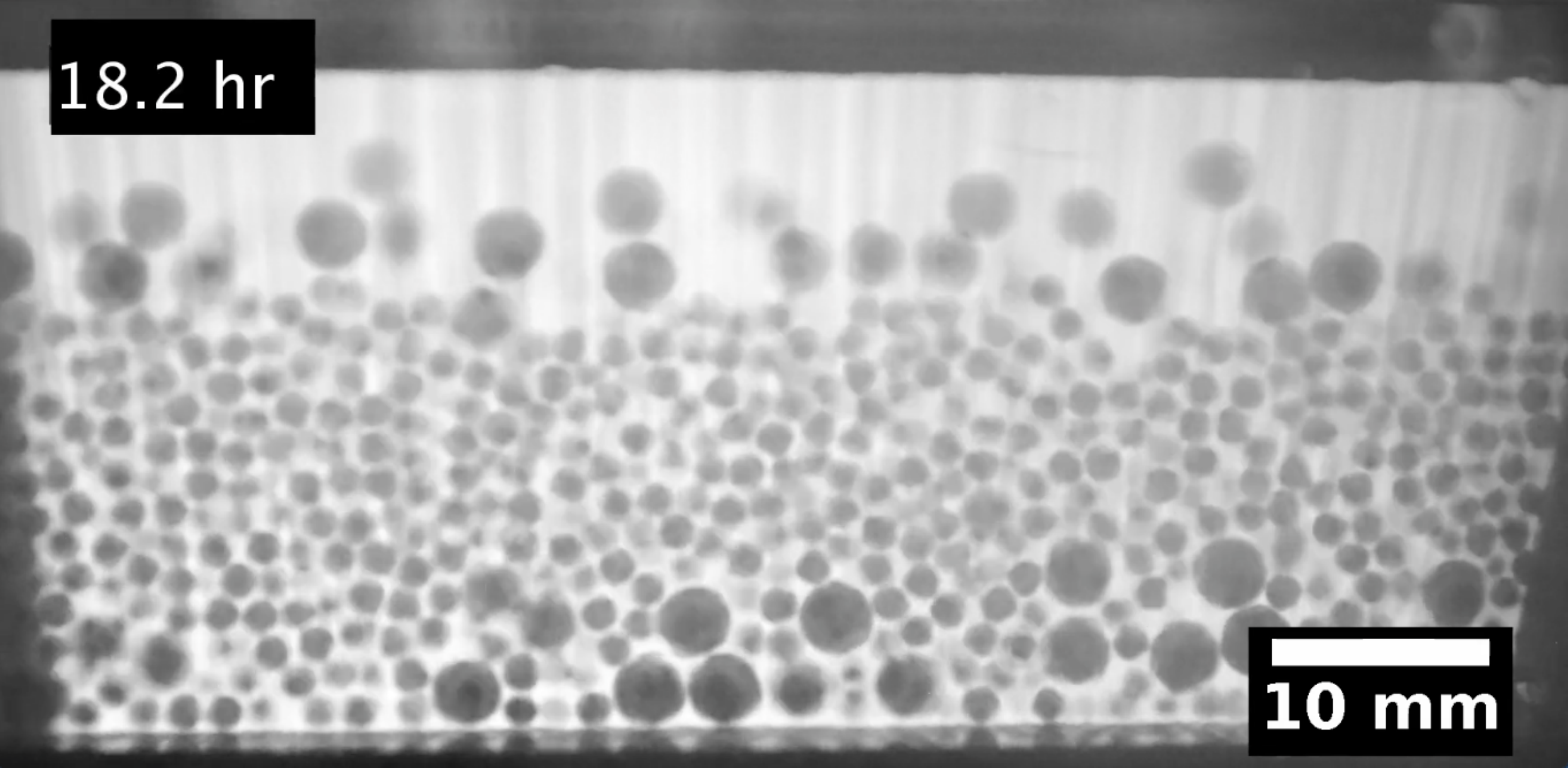}
      \caption{Time-lapse video of the armoring experimental run at shear stress $\tau_s^* = 3.8\tau^*_{cs}$. The real duration of the video is 22.9 hrs, but its playback time is 8 seconds. Note that the snapshots are logarithmically spaced in time.\\
\href{https://www.dropbox.com/s/4x844uurydrk7bv/bidisperse_timelapse.mp4?dl=0}{Movie S2}}
  \label{mov2}
  \end{figure*}

\bibliographystyle{unsrt}
\bibliography{library_segregation}

\end{document}